# Optimal portfolio model based on WVAR

Tianyu Hao

*Master in Financial Engineering, Viterbi School of Engineering*

*University of Southern California*

*Los Angeles, CA, USA*

## 1 Introduction

Portfolio theory is the fundamental of modern finance. In portfolio investment, a crux is how to construct the portfolio. In 1952, H. Markowitz [1] proposed a simple mean - variance approach (1);

The mean-variance approach is to get the result of the optimized problem:

$$\begin{cases} \max \sum_{i=1}^{n} x_i E(r_i) \\ min Var\left(\sum_{i=1}^{n} x_i r_i\right) \\ s.t. \sum_{i=1}^{n} x_i = 1 \\ \quad s.t. x_i \geq 0 \end{cases} \quad (1)$$

Here, variance is used to measure the volatility of asset, especially to measure the risk and volatility of the return of the asset with a distribution of normal distribution. But when used to measure the asset with thick-tail and negative skewness, we may get a wrong estimate of risk, and underestimate the risk. And as most assets don't hold the property of normal distribution, and have thick-tail, so we are seeking for a new measurement of risk.

Value at risk is the central concept of risk management. It determines the maximum amount that a portfolio's value could lose over a given period of time with a given probability, as a result of changes in market prices or rates of return. The denotation of VaR value is

$$q_\lambda(x) = \inf\{x : P(X \leq x) > \lambda\}$$

In the paper of Kostas Giannopoulos[2], VaR method in portfolio selection is compared with traditional mean-variance approach. VaR approach is more intrinsic than variance.

However, VaR don't hold the property of subadditivity, it is not convenient to be used as a optimal goal or constraint in constructing a portfolio. Several examples have been proposed in Rene Garcia's[3] paper that when the tails of two assets are not similar, the VaR of their linear combination might exceed the linear combination of their VaR. Thus $VaR_p(\sum_{i=1}^{n} X_i) > \sum_{i=1}^{n} VaR(X_i)$ .So $\sum_{i=1}^{n} VaR(X_i)$ can't be regarded as the upper bound of $VaR_p(\sum_{i=1}^{n} X_i)$.

Thus we propose a new measurement of risk, coherent risk measure. The theory of coherent risk measure is a very new, important, and rapidly evolving branch of the modern financial mathematics. This concept was introduced by Artzner [4],[5]. Since then, many papers on the topic have followed; surveys of the modern state of the theory are given in [6],[7;Ch.4], and [8]. In some sources theory of coherent risk measures and related topics is already called the "third revolution in finance"(see [9]).

A very important class of coherent risk measures is given by *Tail V@R*. *Tail V@R* of order $\lambda \in [0,1]$ is a map $\rho_\lambda : L^\infty \rightarrow R$ (we have a fixed probability space $(\Omega, F, P)$) defined by $\rho_\lambda(X) = -\inf E_Q X$, where $D_\lambda$ is the set of probability measures $Q$ that are absolutely continuous with respect to P with $\dfrac{dQ}{dP} \leq \lambda^{-1}$. The importance of *Tail V@R* is seen from a result of Kusuoka [10], who proved that $\mu_\lambda$ is the smallest law invariant coherent risk measure that dominates $V@R_\lambda$. This suggests an opinion that *Tail V@R* is one of the best coherent risk measures. For more information on *Tail V@R*, see [9], [10;Sect.6], [11;Sect.7], [12;Sect.4.4]. Here, the integration form of Tail-VAR is

$$\rho_\lambda(X) = \frac{1}{\lambda} \int_0^\lambda q_s(X) ds$$

Here, we can regard Tail VaR as a measurement of the expected loses of an asset or portfolio at the tail of its distribution. In other words, if we know that an asset is going to lose more money than $q_\lambda(X)$, then the expected amount of money the asset will lose is $\rho_\lambda(X)$, the value of its Tail-VaR.

However, there exists a risk measure, which is, in our opinion, much better than *Tail V@R*. It is given by
$$\rho_\mu(X) = \int_{[0,1]} \rho_\lambda(X) \mu(d\lambda), \qquad (2)$$

where $\mu$ is a probability measure on $[0,1]$. We call this measure Weighted Value at Risk and its study is the goal of this paper. Here, the Weighted VaR is just the weighted average of Tail-VaR under the probability measure of $\mu$.

First of all, let us give two arguments in favor of *Weighted V@R* over *Tail V@R*:

- (Financial argument) If the right endpoint of supp $\mu$ is 1, then $\rho_\mu$ depends on the whole distribution of $X$; but *Tail V@R* of order $\lambda$ takes into consideration only $\lambda$-tail of the distribution of $X$.
- (Mathematic argument) If the weighting measure $\mu$ satisfies the condition supp $\mu = [0,1]$, then $\rho_\mu$ possesses some nice properties that are not shared by $\rho_\lambda$. In particular, various optimization problems have a unique solution.

The paper [11] provides some further financial argument in favor of *Weighted V@R*. The first appearance of $\rho_\mu$ in the framework of coherent measures is in the paper of Kusuoka [10]. He proved that any law invariant co monotonic coherent risk measure is of this form. Some further considerations of $\rho_\mu$ can be found in the papers of Acerbi [14], [15], who uses the term spectral risk measure for this class.

Therefore, in consideration of the VAR widely used in the financial field, it is necessary to carry out empirical analysis, and the corresponding numerical method of WVAR, which has superior properties. So far, a few of research literature on TVAR have appeared, but still few numerical solution of the research literature could be found, in order to bring WVAR into practice, this paper will initiate from the TVAR solving problem, based on MATLAB software, using the historical simulation method (avoiding income distribution will be assumed to be normal), the results of previous studies also based on, study the U.S. Nasdaq composite index, combining the Simpson formula for the solution of TVAR and its deeply study; then, through the representation of WVAR formula discussed and indispensable analysis, also using the Simpson formula and the numerical calculations, we have done the empirical analysis and review test.

Furthermore, study of investment risk on the value of VAR measurement and portfolio optimization are becoming increasingly sophisticated (see [16]), this paper is based on WVAR which possesses better properties, taking the idea of portfolio into the multi-index comprehensive evaluation, to build innovative WVAR based portfolio selection under the framework of a theoretical model; in this framework, a description of risks is designed by WVAR, its advantage is no influence by income distribution, meanwhile various optimization problems have a unique solution; then take AHP weights to different indicators deal on this basis, after that we put a nonlinear satisfaction portfolio selected model forward and conduct tests of empirical analysis, finally we use weighted linear approach to convert the portfolio model into a single-objective problem, which is easier to solve, then we use the data of two ETFs to construct portfolio, and compare the performance of portfolio constructed by Mean-Weighted V@R and by Mean-Variance.

## 2 Calculation of WVAR

When calculating the WVAR, this paper considers the situation that the probability measure $\mu = 1$ in order to calculate, combining with the Simpson formula, the specific steps are as follows:

Step1. If we divide integral interval [0,1] into $n = 2m$ deciles, step length $h = \dfrac{b-a}{n}$, node $x_k = a + kh$ $(k = 0,1,...,n)$

Step2. In each segment $[x_{2k-2}, x_{2k}]$ using Simpson formula

$$\int_0^1 f(x)dx \approx \frac{1}{6}[f(1) + 4f(\frac{1}{2}) + f(0)]$$

where $f(x) = TV@R_x(X)$, thus

$$\int_{x_{2k-2}}^{2k} f(x)dx \approx \frac{x_{2k} - x_{2k-2}}{6}[f(x_{2k-2}) + 4f(x_{2k-1}) + f(x_{2k})]$$

$$= \frac{h}{3}[f(x_{2k-2}) + 4f(x_{2k-1}) + f(x_{2k})]$$

where $h = \dfrac{b-a}{n} = \dfrac{x_{2k} - x_{2k-2}}{2}$, sum to it

$$\int_0^1 f(x)dx = \sum_{k=1}^{m} \int_{x_{2k-2}}^{x_{2k}} f(x)dx$$

$$\approx \sum_{k=1}^{m} \frac{h}{3}[f(x_{2k-2}) + 4f(x_{2k-1}) + f(x_{2k})]$$

$$= \frac{h}{3}[\sum_{k=0}^{m-1} f(x_{2k}) + 4\sum_{k=1}^{m} f(x_{2k-1}) + \sum_{k=1}^{m} f(x_{2k})]$$

$$= \frac{h}{3}[f(1) + f(0) + 4\sum_{k=1}^{m} f(x_{2k-1}) + 2\sum_{k=1}^{m-1} f(x_{2k})]$$

Step3. Get the rehabilitation of Simpson formula

$$WV@R(X) = \sum_{k=0}^{n-1} \frac{h}{6}[f(x_k) + 4f(x_{k+\frac{1}{2}}) + f(x_{k+1})]$$

$$= \frac{h}{6}[f(0) + 4\sum_{k=0}^{n-1} f(x_{k+\frac{1}{2}}) + 2\sum_{k=1}^{n-1} f(x_k) + f(1)]$$

## 3 Review test

Based on MATLAB programming platform, according to the Simpson formula above, we have studied the numerical solution of VAR, TVAR and WVAR on the U.S. Nasdaq index, the results are as the following table:

**Table 3.1 Review test**

| Risk Measures | *V@R* | *Tail V@R* | *Weighted V@R* |
|---|---|---|---|
| The overall number of tests | 253 | 253 | 253 |
| The number of test failures | 9 | 2 | 22 |
| Test failure rate | 3.56% | 0.79% | 8.7% |

| Test Conclusion | Risk estimated high | Risk estimated higher | Risk estimated low |
| --- | --- | --- | --- |

As the table shows, through empirical analysis, we can obtain VAR, TVAR and WVAR value of confidence level is 95% before December 31,2009, sum to 253; once these values compared with the actual decline in the stock market respectively, it would arrive at three test derived conclusions, failure rate of WVAR test was 8.7%, which leads to lower risk estimated, in that case the company needs to prepare less risk reserve, for returns more volatile situation it is not conducive to the company during the long investing period; even so, this result is very consistent with WVAR definition of risk for the mean of the entire distribution range, values obtained of WVAR naturally deal with less than VAR, when the volatility of stock market when is flat, then the WVAR in a large extent can ensure the validity of value at risk, meanwhile ensuring more funds to continue investment, which is significant in the financial investment.

## 4 Portfolio Selection Model based on WV@R

With the universal application of VAR in the financial institutions, especially the regulatory authorities have begun to choose VAR to determine the risk capital, a growing number of financial institutions in the investment options consider not only the size of profit margins, but also to meet the constraint of VAR. The situation is that a VAR is given to determine the maximum income portfolio and the constraint.

But the use of VAR as the size of the target to judge the risk can only predict the future risk of short-term value, besides which can't simulate the complex and volatile financial environment, so in the field of portfolio, we may introduce WVAR as a new indicators to characterize the risk measure of market, and establish WVAR-based portfolio selection model.

Unlike a single investment, portfolio risk assessment is decided not only by the risk of each investment form, but also depending on the correlation coefficient between investments and its share size in total investment proportion.

For the given loss probability level $\alpha \in (0,1]$ and the given portfolio $x \in X$, we can obtain risk indicators for each investment according to calculating theories of WVAR above, in order to determine a portfolio comprehensive risk, we introduce the weight coefficient $k_1, k_2 \ldots \ldots k_n$, the weight value of a major investment x can be decided by several factors, such as preference of decision makers and the proportion of portfolio x in the total investment, we can obtain the weight coefficients in use of the expert decision-making method or AHP. There is the overall risk value of W-WVAR (Whole-WVAR) defined as

$$W - WV@R = \sum_{i=1}^{n} k_i WV@R_i \cdot x_i$$

As a matter of fact, W-WV@R can be regarded as the upper bound of the WV@R of a portfolio，as it satisfies $WV@R(\sum k_i \cdot x_i) \leq \sum_{i=1}^{n} k_i WV@R_i \cdot x_i = W - WV@R$

This is determined by the property of Coherent measure of risk.

Artzner et al.(1999) has came up with the idea of coherent measure of risk, he thinks that Coherent measure of risk has these properties

1. Monotonicity：$X \in V, Y \in V, X \leq Y \Rightarrow \rho(X) \geq \rho(Y)$
2. Subadditivity $X \in V, Y \in V, X + Y \in V \Rightarrow \rho(X + Y) \leq \rho(X) + \rho(Y)$
3. Positive homogeneity $X \in V, h > 0, hX \in V \Rightarrow \rho(hX) = h\rho(X)$
4. Translation invariance：$X \in V, a \in R \quad \Rightarrow \quad \rho(X + a) = \rho(X) - a$

The property of subadditivity has guaranteed that the risk of a portfolio is smaller than the sum of risk of each component of the portfolio, which is coincide with the fact that we can reduce non-systematic by diversify investments, and it's a very important property of a risk measurement model.

Kusuoka [12]has proved that $TailV@R$ is a coherent measure of risk，here we will give the proof that WV@R have the property of Coherent measure of risk

Theorem 1. Assume that $\rho_i, i = 1,2,...,n$ is a series of coherent risk measure$a_i, i = 1,2,...,n$ is a series of constant numbers which satisfy$\sum_{i=1}^{n} a_i = 1$, then $\rho = \sum_{i=1}^{n} a_i \rho_i$ belongs to coherent risk measure.

Proof. As $\rho_i$ is coherent risk measure, and for any
$X \in V, Y \in V, X + Y \in V \Rightarrow \rho_i(X + Y) \leq \rho_i(X) + \rho_i(Y)$

$$\rho(X + Y) = \sum_{i=1}^{n} a_i \rho_i(X + Y) \leq \sum_{i=1}^{n} a_i[\rho_i(X) + \rho_i(Y)] = \sum_{i=1}^{n} a_i \rho_i(X) + \sum_{i=1}^{n} a_i \rho_i(Y) = \rho(X) + \rho(Y)$$

As $\rho_i$ is coherent risk measure, so $\rho$ satisfy sub-additivity，similarly we can prove the other three properties.

Theorem 1 has given the result of situation when $\rho_i$ is discrete, and for the situation when $\rho_\alpha$ is continuous, we will see similar results.

Theorem 2 .If $\rho_\alpha$ is continuous coherent risk measure and satisfy $\alpha \in [0,1]$，then for any $du(\alpha)$ which satisfy $\int_0^1 du(\alpha) = 1$，$du(\alpha) \geq 0$，we have the result that $\rho = \int_0^1 \rho_\alpha du(\alpha)$ belongs to coherent risk measurement.

Proof：It's the continuous form of theorem 1. As $\rho_\alpha$ is coherent risk measure, and for any

$X \in V, Y \in V, X + Y \in V$ we have $\rho_\alpha(X+Y) \leq \rho_\alpha(X) + \rho_\alpha(Y)$。And as $du(\alpha)$ satisfy $\int_0^1 du(\alpha) = 1$,

$du(\alpha) \geq 0$，so

$$\rho(X+Y) = \int_0^1 \rho_\alpha(X+Y)du(\alpha) \leq \int_0^1 \left(\rho_\alpha(X) + \rho_\alpha(Y)\right)du(\alpha)$$

$$= \int_0^1 \rho_\alpha(X)du(\alpha) + \int_0^1 \rho_\alpha(Y)du(\alpha) = \rho(X) + \rho(Y)$$

So $\rho$ satisfy subadditivity，besides, as $TailV@R$ is also continuous coherent risk measurement，

which is equivalent to $\rho_\mu$，and $WV@R_u(X) = \int_{[0,1]} TVAR_\lambda(X)ud\lambda$，so $WV@R$ satisfy the property

as well，similarly we can prove the other three properties.

The revenue size of each investment can yield recently proved that according to Markowitz's portfolio theory, a particular rate of return of investors' investing on securities during a period is:

$$R_i = \frac{P_{i1} - P_{i0}}{P_{i0}}$$

Where $R_i$ is return rate of investment for $i$, $P_{i0}$ is the initial capital value, $P_{i1}$ is the sum of investor proceeds during the investment period and the end capital value.

Similarly assuming that the presence of a revenue based weight coefficient $m_1, m_2 \ldots \ldots m_n$, the size is assessed by the preference of the investor proceeds to determine the calculation of the total revenue $R$:

$$R = \sum_{i=1}^n m_i R_i \cdot x_i$$

Ultimately we obtain a portfolio model based on risk indicators of WVAR:

$$\begin{cases} \max \quad R = \sum_{i=1}^n m_i R_i \cdot x_i \\ \min \quad W - WV@R = \sum_{i=1}^n k_i WV@R_i \cdot x_i \\ s.t. \quad \sum_{i=1}^n x_i = 1 \quad x_i \geq 0 \quad (i = 1,2, \ldots \ldots, n); \quad x \in \mathrm{X} \end{cases}$$

As the model is a multi-objective programming problem, we can obtain a single-objective programming from this problem, thus it can be transformed into two kinds of problems:

1. Linear programing:

As our object is to maximize $R = \sum_{i=1}^n m_i R_i \cdot x_i$ and to minimize $W - WV@R = \sum_{i=1}^n k_i WV@R_i \cdot x_i$, so to maximize $R = \sum_{i=1}^n m_i R_i \cdot x_i$ is equivalent to minimize $-\sum_{i=1}^n m_i R_i \cdot x_i$, so we want to minimize $R = \sum_{i=1}^{n} m_i R_i \cdot x_i$ and $-\sum_{i=1}^{n} m_i R_i \cdot x_i$. Thus we can give the W-WV@R a risk aversion parameter.

So the problem here can be transformed to:

$$
\begin{cases}
\min & -\sum_{i=1}^{n} m_i R_i \cdot x_i + m \sum_{i=1}^{n} k_i WV@R_i x_i \\
s.t. & \sum_{i=1}^{n} x_i = 1 \quad x_i \geq 0 \quad (i = 1,2,\ldots,n); \quad x \in X
\end{cases}
$$

Where m is the risk aversion parameter .

As Linear programming is very easy to solve. We can get the numerical solution of some portfolio samples

2. Weighted Linear Method

$$
\min \quad -(\sum_{i=1}^{n} m_i R_i \cdot x_i)^2 + m(\sum_{i=1}^{n} k_i WV@R_i x_i)^2 + J\left(\sum_{i=1}^{n} x_i - 1\right)^2 \quad (3)
$$

Where J is a very large number as a penalty factor to make sure that $\sum_{i=1}^{n} x_i = 1$

And m is a risk aversion parameter.

We can write it as a vector form:

$$
M = \begin{pmatrix} m_1 R_1 \\ m_2 R_2 \\ . \\ . \\ . \\ m_n R_n \end{pmatrix} \quad X = \begin{pmatrix} X_1 \\ X_2 \\ . \\ . \\ . \\ X_n \end{pmatrix} \quad A = \begin{pmatrix} k_1 WVAR_1 \\ k_2 WVAR_2 \\ . \\ . \\ . \\ k_n WVAR_n \end{pmatrix} \quad c = \begin{pmatrix} 1 \\ 1 \\ . \\ . \\ . \\ 1 \end{pmatrix}
$$

then the problem $\min -(\sum_{i=1}^{n} m_i R_i X_i)^2 + m(\sum_{i=1}^{n} k_i WVAR_i X_i)^2 + J(\sum_{i=1}^{n} X_i - 1)^2$ can be rewritten as

$\min -(X^T M)^2 + m(X^T A)^2 + J(c^T X - 1)^2$

Thus we can differentiate the function $-(X^T M)^2 + m(X^T A)^2 + J(c^T X - 1)^2$ to get its optimal solution.

We have $-2M(M^T X) + 2mA(A^T X) + 2Jc(c^T X - 1) = 0$

Simplify it we have $X = (-MM^T + mAA^T + Jcc^T)^{-1} Jc$

## 5 The Compare of Performance of Portfolios we have got from Mean-Variance Approach and Mean-Weighted Value at Risk Approach

As SPY is the ETF based on Standard&Poor's 500 index, which represents the performance of whole stock market, and AGG is the ETF based on the Index of Bond, which also represents the performance of bonds, we can construct our portfolio with these two ETFs to represent the general case in constructing portfolio. Because AGG has been traded since Oct,2003, we use the data of the two ETFs since Oct 3$^{rd}$,2003 until very recently, Sep 28$^{th}$, 2012.

Because trading stocks and bonds have costs, so for most mutual funds, the proportions of portfolios are adjusted periodically, say, once a month. So we adjust our portfolios according to the two different methods once a month. And we use the data before that month to construct monthly portfolio of each approach. Besides, we have used four different risk aversions for each approach to make the compare more complete.

Here is the complete calculating process of each portfolio:

Mean-Weighted V@R Approach:

1. First of all, we use the daily return to calculate WV@R for each ETF with the method of Simpson formula in Matlab. And we use Excel to calculate the mean of return for each ETF.
2. Then we use the approach of Mean-Weighted V@R to calculate the portfolio (optimal solution of (3)) of each month under different risk aversions with the mean and WV@R of the data before this month.
3. After that we use the portfolio of each month to get the mean and WV@R of the portfolio from Jan,2008 to Sep,2012.

Mean-Variance Approach:

1. First of all, we use the daily return to calculate mean and variance of return for each ETF in Excel. And we use Excel to calculate the mean of return for each ETF.
2. Then we use the Mean-Variance approach to calculate the portfolio (to get the optimal solution of (1)) of each month under different risk aversions with the mean and Variance of the data before this month. Here we just transform (1) to be:

$$min - (\sum_{i=1}^{n} R_i \cdot x_i)^2 + mVar\left(\sum_{i=1}^{n} x_i R_i\right)^2 + J\left(\sum_{i=1}^{n} x_i - 1\right)^2$$

3. After that we use the portfolio of each month to get the mean and WV@R of the portfolio from Jan,2008 to Sep,2012.

Finally we have get the performance of portfolios constructed with two approaches here

| High Risk | Mean-WV@R | Mean-Variance |
|---|---|---|
| Mean | -5.5E-05 | -6E-06 |
| WV@R | 4.39E-04 | 3.60E-04 |

| Middle Risk | Mean-WV@R | Mean-Variance |
|---|---|---|
| Mean | 4.62E-05 | 4.25E-05 |
| WV@R | 3.11E-04 | 3.02E-04 |

| Low Risk | Mean-WV@R | Mean-Variance |
|---|---|---|
| Mean | 9.46E-05 | 8.51E-05 |
| WV@R | 2.73E-04 | 2.46E-04 |

| Very Low Risk | Mean-WV@R | Mean-Variance |
|---|---|---|
| Mean | 0.00012 | 0.000112 |
| WV@R | 1.95E-04 | 2.24E-04 |

We can see that no matter in which kind of risk aversion, the mean of return of the portfolio generated by mean-WV@R is higher than that of Mean-Variance approach. And the WV@R of the portfolios of the two approaches are similar, which means that with the method of Mean-WV@R, we can get a portfolio with the same kind of risk and higher return! That's quite an amazing news to those who are seeking for method of constructing optimal portfolio.

After compare, we have found that, at least in some kind of sense, the approach of Mean-WV@R is better than Mean-Variance approach in constructing portfolio.

**6 Conclusion**

The risk is characterized by the obtained WVAR according to the Simpson formula in this paper, which is based on a number of relevant document, besides we establish a portfolio model based on WVAR, gave its optimal analytical solution, and proved that it's better than regular mean-variance approach in constructing portfolio. The experiment proved that VAR and TVAR would overestimate the risk of loss, but when volatility of the stock market is very gentle it is more applicable to use WVAR. This paper only considers the circumstance of the different investments in separate cases, while it is worth of further study of the model extended to the general correlation between the different situations of investments.